\documentstyle[aps,psfig]{revtex}
\begin{document}

\nopagebreak
\title{\hspace{5.0in}Fermilab-Pub-98-258\vspace{1.0in}
Solitary Waves on a Coasting High-Energy Stored Beam}
\author{~S.~I.~Tzenov and ~P.~L.~Colestock}
\address{{\it Fermi National Accelerator Laboratory }\\
{\it P.~O.~Box 500, Batavia, IL 60510, USA}}
\maketitle

\newlength{\figwid} \newlength{\figlen} \medskip 

\begin{abstract}

In this work we derive evolution equations for the nonlinear behavior of
a coasting beam under the influence of a resonator impedance.  Using a
renormalization group approach we find a set of coupled nonlinear equations for 
the beam density and resonator voltage.  Under certain conditions, these may
be analytically solved yielding solitary wave behavior, even in the presence
of significant dissipation in the resonator.  We find long-lived perturbations,
i.e. droplets, which separate from the beam and decelerate toward a quasi-steady
state, in good agreement with simulation results.

\end{abstract}

\section{Introduction.}

Observations of long-lived wave phenomena have been made in stored
high-energy beams for many years. For the most part, these have been ignored
or avoided as pathological conditions that degraded the performance of the
machine. However, in recent experiments, as well as in simulations,
observations have been made which suggest the occurrence of solitary waves
in high-energy stored beams under certain conditions. Both from the point of
view of scientific curiosity as well as the importance of understanding the
formation of halo in such beams, it is worthwhile to study the physics of
these nonlinear waves.

Of particular interest is the saturated state associated with high-intensity
beams under the influence of wakefields, or in the frequency domain, machine
impedance. In stored beams, especially hadron beams where damping mechanisms
are relatively weak, a tenuous equilibrium may develop between beam heating
due to wake-driven fluctuations and damping from a variety of sources. This
state may well be highly nonlinear and may depend on the interaction of
nonlinear waves in order to determine the final equilibrium state. It is our
interest in this work to elucidate the conditions under which nonlinear
waves may occur on a high-energy stored beam. This will then lay the
groundwork for a future study of the evolution of the beam under the
influence of these nonlinear interactions.

We note that much work has been carried out already on solitary waves,
\cite{goldman}, \cite{thornhill}, \cite{robinson}, and references contained
therein,
including those occurring on a beam under the influence of internal space
charge forces \cite{fedele}, \cite{bisognano}. Our situation is new in that we 
consider the specific form of
a wakefield associated with a high-energy beam, namely when space charge
forces are negligible. This leads to a specific form of a solitary wave in a
dissipative system, one which has received limited attention in the literature
thus far \cite{nicholson}, \cite{pereira1}, \cite{pereira2}. 
We have made both experimental observations and carried out
simulations which show the long-lived behavior of the nonlinear waves even
in this dissipative case. It is our aim to shed light on this case.

In this work we adopt an approach which is commonly employed in fluid
dynamics to arrive at a set of model equations for solitary waves on a
coasting beam under the influence of wakefields. It is based on the
renormalization group (RG) analytical approach, which is akin to an envelope
analysis of the wave phenomena. The method in the form we will use it was
introduced by Goldenfeld \cite{chen} and expanded upon by Kunihiro\cite{kunihiro}. 

In Section II we derive the amplitude equations for a resonator impedance
following the standard renormalization group approach. This results in a
nonlinear set of equations for the wave amplitude and beam density. In
Section III we proceed to find analytic solutions for this set which does
indeed admit solitary waves. In Section IV we give the conclusions of this
study and outline the procedure for applying these results to the study of
the steady-state fluctuations on a stored beam.

\section{Derivation of the Amplitude Equations.}

Our starting point is the system of equations

\[
\frac{\partial f}{\partial T}+v\frac{\partial f}{\partial \theta }+\lambda V%
\frac{\partial f}{\partial v}=0, 
\]

\begin{equation}
\frac{\partial ^2V}{\partial T^2}+2\gamma \frac{\partial V}{\partial T}%
+\omega ^2V=\frac{\partial I}{\partial T},  \label{kinetic}
\end{equation}

\[
I\left( \theta ;T\right) =\int dvvf\left( \theta ,v;T\right) 
\]

\noindent for the longitudinal distribution function $f\left( \theta
,v;T\right) $ of an unbunched beam and the voltage variation per turn $%
V\left( \theta ;T\right) $. To write down the equations (\ref{kinetic}) the
following dimensionless variables

\begin{equation}
T=\omega _st\quad ;\quad v=\frac{\stackrel{.}{\theta }}{\omega _s}=1+\frac{%
k_o\epsilon }{\omega _s}\quad ;\quad \omega =\frac{\omega _R}{\omega _s}%
\quad ;\quad \gamma =\frac \omega {2Q},
\end{equation}

\[
\lambda =\frac{e^2{R}k_o\gamma }\pi 
\]

\noindent have been used, where $\omega _s$ is the angular revolution
frequency of the synchronous particle, $\epsilon $ is the energy error, $%
\omega _R$ is the resonator frequency, $Q$ is the quality factor of the
resonator and ${R}$ is the resonator shunt impedance. Furthermore

\begin{equation}
k_o=-\frac{\eta \omega _s}{\beta _s^2E_s}
\end{equation}

\noindent is the proportionality constant between the frequency deviation
and energy deviation of a non synchronous particle with respect to the
synchronous one, while $\eta =\alpha _M-\gamma _s^{-2}$ $\left( \alpha _M%
\text{ - momentum compaction factor}\right) $ is the phase slip coefficient.
The voltage variation per turn $V$ and the beam current $I$ entering eqs. (%
\ref{kinetic}) have been rescaled as well from their actual values $V_a$ and 
$I_a$ according to the relations

\begin{equation}
V_a=2e\omega _s\gamma {R}V\qquad ;\qquad I_a=e\omega _sI.
\end{equation}

Let us now pass to the hydrodynamic description of the longitudinal beam
motion

\[
\frac{\partial \rho }{\partial T}+\frac \partial {\partial \theta }\left(
\rho u\right) =0, 
\]

\[
\frac{\partial u}{\partial T}+u\frac{\partial u}{\partial \theta }=\lambda V-%
\frac{\sigma _v^2}\rho \frac{\partial \rho }{\partial \theta }, 
\]

\[
\frac{\partial ^2V}{\partial T^2}+2\gamma \frac{\partial V}{\partial T}%
+\omega ^2V=\frac \partial {\partial T}\left( \rho u\right) , 
\]

\noindent where

\begin{equation}
\rho \left( \theta ;T\right) =\int dvf\left( \theta ,v;T\right) \qquad
;\qquad \rho \left( \theta ;T\right) u\left( \theta ;T\right) =\int
dvvf\left( \theta ,v;T\right)
\end{equation}

\begin{equation}
\sigma _v=\frac{\left| k_o\right| \sigma _\epsilon }{\omega _s}
\end{equation}

\noindent and $\sigma _\epsilon $ is the r.m.s. of the energy error that is
proportional to the longitudinal beam temperature. Rescaling further the
variables $\rho $ and $V$ according to

\begin{equation}
\rho _a=\rho _o\rho \qquad ;\qquad V_a=2e\omega _s\rho _o\gamma {R}V\qquad
;\qquad \lambda =\frac{e^2{R}\gamma k_o\rho _o}\pi
\end{equation}

\noindent and taking onto account that the dependence of all hydrodynamic
variables on $\theta $ is slow $\left( \sim \varepsilon \theta \right) $
compared to the dependence on time we write the gas-dynamic equations as

\[
\frac{\partial \rho }{\partial T}+\varepsilon \frac \partial {\partial
\theta }\left( \rho u\right) =0, 
\]

\begin{equation}
\frac{\partial u}{\partial T}+\varepsilon u\frac{\partial u}{\partial \theta 
}=\lambda V-\varepsilon \frac{\sigma _v^2}\rho \frac{\partial \rho }{%
\partial \theta },  \label{hydro}
\end{equation}

\[
\frac{\partial ^2V}{\partial T^2}+2\gamma \frac{\partial V}{\partial T}%
+\omega ^2V=\frac \partial {\partial T}\left( \rho u\right) . 
\]

\noindent Here $\varepsilon $ is a formal perturbation parameter, which is
set to unity at the end of the calculations and should not be confused with
the energy error variable. We will derive slow motion equations from the
system (\ref{hydro}) by means of the renormalization group (RG) approach
\cite{chen}, \cite{kunihiro}.  To do so we perform a naive perturbation expansion

\begin{equation}
\rho =1+\sum\limits_{m=1}^\infty \varepsilon ^m\rho _m\qquad ;\qquad
u=1+\sum\limits_{m=1}^\infty \varepsilon ^mu_m\qquad ;\qquad
V=\sum\limits_{m=1}^\infty \varepsilon ^mV_m
\end{equation}

\noindent around the stationary solution

\begin{equation}
\rho ^{\left( 0\right) }=1\qquad ;\qquad u^{\left( 0\right) }=1\qquad
;\qquad V^{\left( 0\right) }=0.
\end{equation}

\noindent The first order equations are

\[
\frac{\partial \rho _1}{\partial T}=0\qquad ;\qquad \frac{\partial u_1}{%
\partial T}=\lambda V_1\qquad ;\qquad \frac{\partial ^2V_1}{\partial T^2}%
+2\gamma \frac{\partial V_1}{\partial T}+\omega ^2V_1=\frac{\partial u_1}{%
\partial T} 
\]

\noindent with obvious solution

\begin{equation}
V_1\left( \theta ;T\right) =E\left( \theta ;T_o\right) e^{i\omega _1\Delta
T}+E^{*}\left( \theta ;T_o\right) e^{-i\omega _1^{*}\Delta T},  \label{volt1}
\end{equation}

\begin{equation}
u_1\left( \theta ;T\right) =u_o\left( \theta ;T_o\right) +\lambda \left[ 
\frac{E\left( \theta ;T_o\right) }{i\omega _1}e^{i\omega _1\Delta T}-\frac{%
E^{*}\left( \theta ;T_o\right) }{i\omega _1^{*}}e^{-i\omega _1^{*}\Delta
T}\right] ,  \label{veloc1}
\end{equation}

\begin{equation}
\rho _1\left( \theta ;T\right) =R_o\left( \theta ;T_o\right) .  \label{dens1}
\end{equation}

\noindent In expressions (\ref{volt1}-\ref{dens1}) the following notations
have been introduced

\begin{equation}
\omega _1=\omega _q+i\gamma \qquad ;\qquad \omega _q^2=\omega _o^2-\gamma
^2\qquad ;\qquad \omega _o^2=\omega ^2-\lambda ,
\end{equation}

\begin{equation}
\Delta T=T-T_o,
\end{equation}

\noindent where the amplitudes $E\left( \theta ;T_o\right) $, $u_o\left(
\theta ;T_o\right) $, $R_o\left( \theta ;T_o\right) $ are yet unknown
functions of $\theta $ and the initial instant of time $T_o$. Proceeding
further we write down the second order equations

\[
\frac{\partial \rho _2}{\partial T}+\frac \partial {\partial \theta }\left(
\rho _1+u_1\right) =0, 
\]

\[
\frac{\partial u_2}{\partial T}+\frac{\partial u_1}{\partial \theta }%
=\lambda V_2-\sigma _v^2\frac{\partial \rho _1}{\partial \theta }, 
\]

\[
\frac{\partial ^2V_2}{\partial T^2}+2\gamma \frac{\partial V_2}{\partial T}%
+\omega ^2V_2=\frac \partial {\partial T}\left( u_2+\rho _1u_1+\rho
_2\right) . 
\]

\noindent Solving the equation for the voltage

\[
\frac{\partial ^2V_2}{\partial T^2}+2\gamma \frac{\partial V_2}{\partial T}%
+\omega _o^2V_2=-2\frac{\partial u_1}{\partial \theta }-\left( \sigma
_v^2+1\right) \frac{dR_o}{d\theta }+\lambda R_oV_1 
\]

\noindent that can be obtained by combining the second order equations, and
subsequently the other two equations for $u_2$ and $\rho _2$ we find

\begin{equation}
V_2\left( \theta ;T\right) =-\frac 1{\omega _o^2}\left[ 2u_o^{\prime
}+\left( \sigma _v^2+1\right) R_o^{\prime }\right] +\frac{\lambda \Delta T}{%
2i\omega _q}\left( R_oE+\frac{2iE^{\prime }}{\omega _1}\right) e^{i\omega
_1\Delta T}+c.c.
\end{equation}

\[
u_2\left( \theta ;T\right) =-\left\{ u_o^{\prime }+\sigma _v^2R_o^{\prime }+%
\frac \lambda {\omega _o^2}\left[ 2u_o^{\prime }+\left( \sigma _v^2+1\right)
R_o^{\prime }\right] \right\} \Delta T+\frac \lambda {\omega _1^2}E^{\prime
}e^{i\omega _1\Delta T}+ 
\]

\begin{equation}
+\frac{\lambda ^2}{2i\omega _q\omega _1^2}\left( R_oE+\frac{2iE^{\prime }}{%
\omega _1}\right) e^{i\omega _1\Delta T}-\frac{\lambda ^2\Delta T}{2\omega
_q\omega _1}\left( R_oE+\frac{2iE^{\prime }}{\omega _1}\right) e^{i\omega
_1\Delta T}+c.c.
\end{equation}

\begin{equation}
\rho _2\left( \theta ;T\right) =-\left( R_o^{\prime }+u_o^{\prime }\right)
\Delta T+\frac \lambda {\omega _1^2}E^{\prime }e^{i\omega _1\Delta T}+c.c.
\end{equation}

\noindent where the prime implies differentiation with respect to $\theta $.
In a similar way we obtain the third order equations

\[
\frac{\partial \rho _3}{\partial T}+\frac \partial {\partial \theta }\left(
\rho _2+R_ou_1+u_2\right) =0, 
\]

\[
\frac{\partial u_3}{\partial T}+u_1\frac{\partial u_1}{\partial \theta }+%
\frac{\partial u_2}{\partial \theta }=\lambda V_3-\sigma _v^2\left( \frac{%
\partial \rho _2}{\partial \theta }-R_o\frac{dR_o}{d\theta }\right) , 
\]

\[
\frac{\partial ^2V_3}{\partial T^2}+2\gamma \frac{\partial V_3}{\partial T}%
+\omega ^2V_3=\frac \partial {\partial T}\left( u_3+R_ou_2+\rho _2u_1+\rho
_3\right) . 
\]

\noindent Solving the equation for the voltage

\[
\frac{\partial ^2V_3}{\partial T^2}+2\gamma \frac{\partial V_3}{\partial T}%
+\omega _o^2V_3=-2u_2^{\prime }-2u_1u_1^{\prime }-\left( \sigma
_v^2+1\right) \rho _2^{\prime }+\lambda R_oV_2-2\left( R_ou_1\right)
^{\prime }+\lambda \rho _2V_1 
\]

\noindent that can be obtained by combining the third order equations, and
subsequently the other two equations for $u_3$ and $\rho _3$ we obtain

\[
V_3\left( \theta ;T\right) =-\frac 1{\omega _o^2}\left\{ 2u_ou_o^{\prime }+%
\frac \lambda {\omega _o^2}\left[ 2R_ou_o^{\prime }+\left( \sigma
_v^2+1\right) R_oR_o^{\prime }\right] +2\left( R_ou_o\right) ^{\prime
}\right\} 
\]

\[
-\frac{2\gamma }{\omega _o^4}\left\{ 2\left[ u_o^{\prime \prime }+\sigma
_v^2R_o^{\prime \prime }+\frac \lambda {\omega _o^2}\left( 2u_o^{\prime
\prime }+\left( \sigma _v^2+1\right) R_o^{\prime \prime }\right) \right]
+\left( \sigma _v^2+1\right) \left( u_o^{\prime \prime }+R_o^{\prime \prime
}\right) \right\} 
\]

\[
+\frac{\Delta T}{\omega _o^2}\left\{ 2\left[ u_o^{\prime \prime }+\sigma
_v^2R_o^{\prime \prime }+\frac \lambda {\omega _o^2}\left( 2u_o^{\prime
\prime }+\left( \sigma _v^2+1\right) R_o^{\prime \prime }\right) \right]
+\left( \sigma _v^2+1\right) \left( u_o^{\prime \prime }+R_o^{\prime \prime
}\right) \right\} 
\]

\[
+\frac 1{\omega _o^2}\left\{ -\frac{2\lambda ^2}{\omega _o^2}\left( \left|
E\right| ^2\right) ^{\prime }+\lambda ^2\left( \frac{E^{\prime }E^{*}}{%
\omega _1^2}+\frac{EE^{*\prime }}{\omega _1^{*2}}\right) \right\}
e^{-2\gamma \Delta T}+ 
\]

\[
\frac{\lambda \Delta T}{2i\omega _q}\left\{ -\left( \sigma _v^2+3\right) 
\frac{E^{\prime \prime }}{\omega _1^2}+\frac \lambda {\omega _q\omega _1^2}%
\left[ i\left( R_oE\right) ^{\prime }-\frac{2E^{\prime \prime }}{\omega _1}%
\right] +\frac{2i}{\omega _1}\left[ \left( u_o+R_o\right) E\right] ^{\prime
}\right\} e^{i\omega _1\Delta T} 
\]

\[
+\frac{\lambda \Delta T}{4\omega _q^2}\left( 1-i\omega _q\Delta T\right) * 
\]

\begin{equation}
\ast \left\{ \frac \lambda {\omega _1\omega _q}\left[ \left( R_oE\right)
^{\prime }+\frac{2iE^{\prime \prime }}{\omega _1}\right] +\frac{\lambda R_o}{%
2i\omega _q}\left( R_oE+\frac{2iE^{\prime }}{\omega _1}\right) -\left(
u_o^{\prime }+R_o^{\prime }\right) E\right\} e^{i\omega _1\Delta T}+c.c.
\end{equation}

\[
u_3\left( \theta ;T\right) =\left( \sigma _v^2R_oR_o^{\prime
}-u_ou_o^{\prime }\right) \Delta T+\frac{\lambda ^2}{2\gamma \omega _o^2}%
\frac{\partial \left| E\right| ^2}{\partial \theta }e^{-2\gamma \Delta T}- 
\]

\[
-\frac \lambda {\omega _o^2}\left\{ 2u_ou_o^{\prime }+\frac{\lambda R_o}{%
\omega _o^2}\left[ 2u_o^{\prime }+\left( \sigma _v^2+1\right) R_o^{\prime
}\right] +2\left( R_ou_o\right) ^{\prime }\right\} \Delta T- 
\]

\[
-\frac{2\gamma \lambda }{\omega _o^4}\left\{ 2\left[ u_o^{\prime \prime
}+\sigma _v^2R_o^{\prime \prime }+\frac \lambda {\omega _o^2}\left(
2u_o^{\prime \prime }+\left( \sigma _v^2+1\right) R_o^{\prime \prime
}\right) \right] +\left( \sigma _v^2+1\right) \left( u_o^{\prime \prime
}+R_o^{\prime \prime }\right) \right\} \Delta T 
\]

\[
-\frac \lambda {2\gamma \omega _o^2}\left\{ -\frac{2\lambda ^2}{\omega _o^2}%
\frac{\partial \left| E\right| ^2}{\partial \theta }+\lambda ^2\left( \frac{%
E^{\prime }E^{*}}{\omega _1^2}+\frac{EE^{*\prime }}{\omega _1^{*2}}\right)
\right\} e^{-2\gamma \Delta T}+ 
\]

\begin{equation}
+\text{ oscillating terms and terms proportional to }\left( \Delta T\right)
^2,
\end{equation}

\[
\rho _3\left( \theta ;T\right) =-\left( R_ou_o\right) ^{\prime }T+ 
\]

\begin{equation}
+\text{ oscillating terms and terms proportional to }\left( \Delta T\right)
^2.
\end{equation}

\noindent Collecting most singular terms that would contribute to the
amplitude equations when applying the RG procedure, and setting $\varepsilon
=1$ we write down the following expressions for $V_{RG}$, $u_{RG}$ and $\rho
_{RG}$

\[
V_{RG}\left( \theta ;T,T_o\right) =Ee^{i\omega _1\Delta T}+ 
\]

\[
\frac{\lambda \Delta T}{2i\omega _q}\left\{ R_oE+\frac{2i}{\omega _1}\left[
\left( 1+u_o+R_o\right) E\right] ^{\prime }-\left( \sigma _v^2+3\right) 
\frac{E^{\prime \prime }}{\omega _1^2}-\frac i{2\omega _q}\left(
u_o+R_o\right) ^{\prime }E\right\} e^{i\omega _1\Delta T} 
\]

\begin{equation}
+\frac{\lambda ^2\Delta T}{2i\omega _q^2}\left\{ \frac{\omega _1+2\omega _q}{%
2\omega _1^2\omega _q}\left[ i\left( R_oE\right) ^{\prime }-\frac{2E^{\prime
\prime }}{\omega _1}\right] +\frac{R_o}{4\omega _q}\left( R_oE+\frac{%
2iE^{\prime }}{\omega _1}\right) \right\} e^{i\omega _1\Delta T}+c.c.
\label{pert1}
\end{equation}

\[
u_{RG}\left( \theta ;T\right) =u_o-\left\{ u_o^{\prime }+\sigma
_v^2R_o^{\prime }+\frac \lambda {\omega _o^2}\left[ 2u_o^{\prime }+\left(
\sigma _v^2+1\right) R_o^{\prime }\right] \right\} \Delta T+ 
\]

\[
+\left( \sigma _v^2R_oR_o^{\prime }-u_ou_o^{\prime }\right) \Delta T+\frac{%
\lambda ^2}{2\gamma \omega _o^2}\frac{\partial \left| E\right| ^2}{\partial
\theta }e^{-2\gamma \Delta T}- 
\]

\[
-\frac{\lambda \Delta T}{\omega _o^2}\left\{ 2u_ou_o^{\prime }+\frac{\lambda
R_o}{\omega _o^2}\left[ 2u_o^{\prime }+\left( \sigma _v^2+1\right)
R_o^{\prime }\right] +2\left( R_ou_o\right) ^{\prime }\right\} - 
\]

\[
-\frac{2\gamma \lambda \Delta T}{\omega _o^4}\left\{ 2\left[ u_o^{\prime
\prime }+\sigma _v^2R_o^{\prime \prime }+\frac \lambda {\omega _o^2}\left(
2u_o^{\prime \prime }+\left( \sigma _v^2+1\right) R_o^{\prime \prime
}\right) \right] +\left( \sigma _v^2+1\right) \left( u_o^{\prime \prime
}+R_o^{\prime \prime }\right) \right\} 
\]

\begin{equation}
-\frac \lambda {2\gamma \omega _o^2}\left\{ -\frac{2\lambda ^2}{\omega _o^2}%
\frac{\partial \left| E\right| ^2}{\partial \theta }+\lambda ^2\left( \frac{%
E^{\prime }E^{*}}{\omega _1^2}+\frac{EE^{*\prime }}{\omega _1^{*2}}\right)
\right\} e^{-2\gamma \Delta T},  \label{pert2}
\end{equation}

\begin{equation}
\rho _{RG}\left( \theta ;T\right) =R_o-\left[ R_o^{\prime }+u_o^{\prime
}+\left( R_ou_o\right) ^{\prime }\right] \Delta T.  \label{pert3}
\end{equation}

The amplitudes $E$, $u_o$ and $R_o$ can be renormalized so as to remove the
secular terms in the above expressions (\ref{pert1}-\ref{pert3}) and thus
obtain the corresponding RG equations. Not entering into details let us
briefly state the basic features of the RG approach\cite{kunihiro}.  The perturbative
solution (\ref{pert1}-\ref{pert3}) can be regarded as a parameterization of
a 3D family of curves $\left\{ \Re _{T_o}\right\} =\left( R_o\left(
T_o\right) ,\ u_o\left( T_o\right) ,\ E\left( T_o\right) \right) $ with $T_o$
being a free parameter. It can be shown that the RG equations are precisely
the envelope equations for the one -parameter family $\left\{ \Re
_{T_o}\right\} :$

\begin{equation}
\left. \left( \frac{\partial R_o}{\partial T_o},\ \frac{\partial u_o}{%
\partial T_o},\ \frac{\partial E}{\partial T_o}\right) \right| _{T_o=T}=0.
\end{equation}

\noindent It is straightforward now to write down the RG equations in our
case as follows:

\begin{equation}
\frac{\partial R_o}{\partial T}+\frac \partial {\partial \theta }\left(
R_o+u_o+R_ou_o\right) =0,  \label{densenv}
\end{equation}

\[
\frac{\partial u_o}{\partial T}+\frac \partial {\partial \theta }\left(
u_o+\sigma _v^2R_o\right) +u_o\frac{\partial u_o}{\partial \theta }-\sigma
_v^2R_o\frac{\partial R_o}{\partial \theta }+\frac{\lambda ^2}{\omega _o^2}%
\frac{\partial \left| E\right| ^2}{\partial \theta }e^{-2\gamma T}= 
\]

\[
=-\frac{2\lambda }{\omega _o^2}\left[ u_o^{\prime }+u_ou_o^{\prime }+\left(
R_ou_o\right) ^{\prime }+\left( \sigma _v^2+1\right) \frac{R_o^{\prime }}2%
\right] -\frac{\lambda ^2R_o}{\omega _o^4}\left[ 2u_o^{\prime }+\left(
\sigma _v^2+1\right) R_o^{\prime }\right] - 
\]

\[
-\frac{2\gamma \lambda }{\omega _o^4}\left\{ 2u_o^{\prime \prime }+2\sigma
_v^2R_o^{\prime \prime }+\frac{2\lambda }{\omega _o^2}\left[ 2u_o^{\prime
\prime }+\left( \sigma _v^2+1\right) R_o^{\prime \prime }\right] +\left(
\sigma _v^2+1\right) \left( R_o^{\prime \prime }+u_o^{\prime \prime }\right)
\right\} - 
\]

\begin{equation}
-\frac \lambda {\omega _o^2}\left[ \frac{2\lambda ^2}{\omega _o^2}\frac{%
\partial \left| E\right| ^2}{\partial \theta }-\lambda ^2\left( \frac{%
E^{\prime }E^{*}}{\omega _1^2}+\frac{EE^{*\prime }}{\omega _1^{*2}}\right)
\right] e^{-2\gamma T},  \label{velenv}
\end{equation}

\[
\frac{2i\omega _q}\lambda \left( \frac \partial {\partial T}+\frac \partial {%
\partial \theta }\right) E=R_oE+\frac{2i}{\omega _1}\left[ \left(
1+u_o+R_o\right) E\right] ^{\prime }-\frac{\sigma _v^2+3}{\omega _1^2}%
E^{\prime \prime }- 
\]

\[
-\frac i{2\omega _q}\left( 1+u_o+R_o\right) ^{\prime }E+ 
\]

\begin{equation}
+\frac \lambda {\omega _q}\left\{ \frac{\omega _1+2\omega _q}{2\omega
_1^2\omega _q}\left[ i\left( R_oE\right) ^{\prime }-\frac{2E^{\prime \prime }%
}{\omega _1}\right] +\frac{R_o}{4\omega _q}\left( R_oE+\frac{2iE^{\prime }}{%
\omega _1}\right) \right\} .  \label{voltenv}
\end{equation}

\noindent In deriving eq. (\ref{voltenv}) we have assumed that the voltage
envelope function $E$ depends on its arguments as $E\left( \theta
-T_o;T_o\right) $. Neglecting higher order terms we finally obtain the
desired equations governing the evolution of the amplitudes

\begin{equation}
\frac{\partial \widetilde{\rho }}{\partial T}+\frac \partial {\partial
\theta }\left( \widetilde{\rho }\widetilde{u}\right) =0,  \label{dense}
\end{equation}

\begin{equation}
\frac{\partial \widetilde{u}}{\partial T}+\widetilde{u}\frac{\partial 
\widetilde{u}}{\partial \theta }=-\frac{\sigma _v^2}{\widetilde{\rho }}\frac{%
\partial \widetilde{\rho }}{\partial \theta }-\frac{\lambda ^2}{\omega _o^2}%
\frac{\partial \left| \widetilde{E}\right| ^2}{\partial \theta },
\label{vele}
\end{equation}

\[
\frac{2i\omega _q}\lambda \left( \frac \partial {\partial T}+\frac \partial {%
\partial \theta }+\gamma \right) \widetilde{E}=\left( \widetilde{\rho }%
-1\right) \widetilde{E}-\frac{\sigma _v^2+3}{\omega _1^2}\frac{\partial ^2%
\widetilde{E}}{\partial \theta ^2}+ 
\]

\begin{equation}
+\frac{2i}{\omega _1}\frac \partial {\partial \theta }\left( \widetilde{\rho 
}\widetilde{u}\widetilde{E}\right) +\frac i{2\omega _q}\widetilde{E}\frac{%
\partial \widetilde{\rho }}{\partial T},  \label{volte}
\end{equation}

\noindent where

\begin{equation}
\widetilde{\rho }=1+R_o\qquad ;\qquad \widetilde{u}=1+u_o\qquad ;\qquad 
\widetilde{E}=Ee^{-\gamma T}.
\end{equation}

\noindent Eliminating $\widetilde{u}$ from equations (\ref{dense}) and (\ref
{vele}) we get

\begin{equation}
\frac{\partial ^2\widetilde{\rho }}{\partial T^2}-\sigma _v^2\frac{\partial
^2\widetilde{\rho }}{\partial \theta ^2}=\frac{\lambda ^2}{\omega _o^2}\frac{%
\partial ^2\left| \widetilde{E}\right| ^2}{\partial \theta ^2},
\label{zakharov1}
\end{equation}

\begin{equation}
\frac{2i\omega _q}\lambda \left( \frac \partial {\partial T}+\frac \partial {%
\partial \theta }+\gamma \right) \widetilde{E}=-\frac{\sigma _v^2+3}{\omega
_1^2}\frac{\partial ^2\widetilde{E}}{\partial \theta ^2}+\frac{2i}{\omega _1}%
\frac{\partial \widetilde{E}}{\partial \theta }+\left( \widetilde{\rho }%
-1\right) \widetilde{E}.  \label{zakharov2}
\end{equation}

\section{Solution of the Amplitude Equations.}

Let us perform a scaling of variables in the amplitude equations (\ref
{zakharov1}), (\ref{zakharov2}) according to the relations

\begin{equation}
\tau =\frac{\lambda T}{2\omega _q}\qquad ;\qquad \Theta =\frac{\omega
_o\theta }{\sqrt{\sigma _v^2+3}}\qquad ;\qquad \psi =\frac{\left| \lambda
\right| \widetilde{E}}{\sigma _v\omega _o}.
\end{equation}

\noindent The amplitude equations take now the form

\begin{equation}
\frac{\partial ^2\widetilde{\rho }}{\partial \tau ^2}-c_u^2\frac{\partial ^2%
\widetilde{\rho }}{\partial \Theta ^2}=c_u^2\frac{\partial ^2\left| \psi
\right| ^2}{\partial \Theta ^2}  \label{scale1}
\end{equation}

\begin{equation}
i\left( \frac \partial {\partial \tau }+\frac{ab\omega _o}2\frac \partial {%
\partial \Theta }\right) \psi +i\gamma b\psi =-\frac{\omega _o^2}{\omega _1^2%
}\frac{\partial ^2\psi }{\partial \Theta ^2}+ia\frac{\omega _o}{\omega _1}%
\frac{\partial \psi }{\partial \Theta }+\left( \widetilde{\rho }-1\right)
\psi ,  \label{scale2}
\end{equation}

\noindent where

\begin{equation}
a=\frac 2{\sqrt{\sigma _v^2+3}}\qquad ;\qquad b=\frac{2\omega _q}\lambda
\qquad ;\qquad c_u=\frac{2\sigma _v\omega _q\omega _o}{\left| \lambda
\right| \sqrt{\sigma _v^2+3}}.
\end{equation}

\noindent From equation (\ref{scale1}) one finds approximately

\begin{equation}
\widetilde{\rho }=1-\left| \psi \right| ^2.  \label{caviton}
\end{equation}

\noindent Equation (\ref{caviton}) when substituted into (\ref{scale2})
yields

\begin{equation}
i\left( \frac \partial {\partial \tau }+\frac{ab\omega _o}2\frac \partial {%
\partial \Theta }\right) \psi +i\gamma b\psi =-\frac{\omega _o^2}{\omega _1^2%
}\frac{\partial ^2\psi }{\partial \Theta ^2}+ia\frac{\omega _o}{\omega _1}%
\frac{\partial \psi }{\partial \Theta }-\left| \psi \right| ^2\psi .
\label{schrod}
\end{equation}

\noindent Noting that

\[
\omega _1=\omega _oe^{i\omega _{\arg }}\qquad \quad ;\qquad \quad \omega
_{\arg }=\arctan \frac \gamma {\omega _q} 
\]

\noindent and introducing the new variable

\begin{equation}
x=\Theta +a\tau -\frac{ab\omega _o}2\tau =\frac{\omega _o}{\sqrt{\sigma
_v^2+3}}\left( \theta -T+\frac{\lambda T}{\omega _o\omega _q}\right)
\end{equation}

\noindent we rewrite the nonlinear Schr\"odinger equation (\ref{schrod}) in
the form \cite{pereira1}

\begin{equation}
i\frac{\partial \psi }{\partial \tau }+i\gamma b\psi =-\left( 1-\frac{%
2i\gamma }{\omega _o}\right) \frac{\partial ^2\psi }{\partial x^2}+\frac{%
a\gamma }{\omega _o}\frac{\partial \psi }{\partial x}-\left| \psi \right|
^2\psi .  \label{nlse}
\end{equation}

Next we examine the linear stability of the solution

\begin{equation}
\psi _o\left( x;\tau \right) =A_oe^{i\left( kx-\Omega \tau \right) },
\label{solut}
\end{equation}

\noindent where

\[
\Omega =k^2-A_o^2-\frac{i\gamma }{\omega _o}\left( 2k^2-ak+\omega _ob\right)
. 
\]

\noindent In the case the energy of the beam is above transition energy $%
\left( k_o<0\right) $ the solution (\ref{solut}) is exponentially decaying
for

\begin{equation}
1-\sqrt{1+\frac{8\omega _o\left| b\right| }{a^2}}<\frac{4k}a<1+\sqrt{1+\frac{%
8\omega _o\left| b\right| }{a^2}}
\end{equation}

To proceed further let us represent the field envelope function $\psi $ as

\begin{equation}
\psi \left( x;\tau \right) ={A}\left( x;\tau \right) e^{i\varphi \left(
x;\tau \right) }
\end{equation}

\noindent and write the equations for the amplitude ${A}$ and the phase $%
\varphi $

\begin{equation}
{A}_\tau +\gamma b{A}=-{A\varphi }_{xx}-2{A}_x\varphi _x+\frac{2\gamma }{%
\omega _o}\left( {A}_{xx}-{A\varphi }_x^2\right) +\frac{a\gamma }{\omega _o}{%
A}\varphi _x,  \label{ampl}
\end{equation}

\begin{equation}
{A}\varphi _\tau ={A}_{xx}-{A\varphi }_x^2+{A}^3+\frac{2\gamma }{\omega _o}%
\left( {A\varphi }_{xx}+2{A}_x\varphi _x\right) -\frac{a\gamma }{\omega _o}{A%
}_x.  \label{phase}
\end{equation}

\noindent When $\gamma =0$ the above system admits a simple one-soliton
solution of the form

\begin{equation}
\varphi \left( x;\tau \right) =kx-\Omega \tau +\alpha ,
\end{equation}

\begin{equation}
{A}\left( x;\tau \right) =\frac{\sqrt{2}K}{\cosh \left[ K\left( x-2k\tau
+\beta \right) \right] }\qquad ;\qquad K^2=k^2-\Omega >0.
\end{equation}

\noindent Define now the quantities

\begin{equation}
{N}\left( \tau \right) =\int dx\left| \psi \left( x;\tau \right) \right|
^2\qquad ;\qquad {P}\left( \tau \right) =\frac i2\int dx\left( \psi \frac{%
\partial \psi ^{*}}{\partial x}-\psi ^{*}\frac{\partial \psi }{\partial x}%
\right) .
\end{equation}

\noindent These are the first two (particle density and momentum
respectively) from the infinite hierarchy of integrals of motion for the
undamped $\left( \gamma =0\right) $ nonlinear Schr\"odinger equation 
\cite{zakharov}. When
damping is present $\left( \gamma \neq 0\right) $ they are no longer
integrals of motion and their dynamics is governed by the equations

\begin{equation}
\frac{d{N}}{d\tau }+2\gamma b{N}=-\frac{4\gamma }{\omega _o}\int dx\left| 
\frac{\partial \psi }{\partial x}\right| ^2+\frac{2a\gamma }{\omega _o}{P},
\label{integral1}
\end{equation}

\begin{equation}
\frac{d{P}}{d\tau }+2\gamma b{P}=\frac{2i\gamma }{\omega _o}\int dx\left( 
\frac{\partial ^2\psi }{\partial x^2}\frac{\partial \psi ^{*}}{\partial x}-%
\frac{\partial ^2\psi ^{*}}{\partial x^2}\frac{\partial \psi }{\partial x}%
\right) +\frac{2a\gamma }{\omega _o}\int dx\left| \frac{\partial \psi }{%
\partial x}\right| ^2.  \label{integral2}
\end{equation}

\noindent Instead of solving equations (\ref{ampl}) and (\ref{phase}) for
the amplitude ${A}$ and the phase $\varphi $ we approximate the solution of
the nonlinear Schr\"odinger equation (\ref{nlse}) with a one-soliton
travelling wave

\begin{equation}
\psi \left( x;\tau \right) =\frac{\sqrt{2}\eta \left( \tau \right) }{\cosh
\left\{ \eta \left( \tau \right) \left[ x-\mu \left( \tau \right) +\beta
\right] \right\} }\exp \left\{ i\left[ \sigma \left( \tau \right) x-\Omega
\left( \tau \right) +\alpha \right] \right\} ,  \label{kink}
\end{equation}

\noindent where

\begin{equation}
\mu \left( \tau \right) =2\int d\tau \sigma \left( \tau \right) \qquad
;\qquad \Omega \left( \tau \right) =\int d\tau \left[ \sigma ^2\left( \tau
\right) -\eta ^2\left( \tau \right) \right] .
\end{equation}

\noindent Substituting the sample solution (\ref{kink}) into the balance
equations (\ref{integral1}), (\ref{integral2}) and noting that

\[
{N}\left( \tau \right) =4\eta \left( \tau \right) \qquad ;\qquad {P}\left(
\tau \right) =4\eta \left( \tau \right) \sigma \left( \tau \right) 
\]

\noindent we obtain the following system of equations

\[
\frac{d\eta }{d\tau }+2\gamma b\eta =-\frac{4\gamma }{\omega _o}\left( \frac{%
\eta ^3}3+\eta \sigma ^2\right) +\frac{2a\gamma }{\omega _o}\eta \sigma , 
\]

\[
\frac{d\left( \eta \sigma \right) }{d\tau }+2\gamma b\eta \sigma =-\frac{%
4\gamma }{\omega _o}\left( \eta ^3\sigma +\eta \sigma ^3\right) +\frac{%
2a\gamma }{\omega _o}\left( \frac{\eta ^3}3+\eta \sigma ^2\right) , 
\]

\noindent or

\begin{equation}
\frac{d\eta }{d\tau }+2\gamma b\eta =-\frac{4\gamma }{\omega _o}\left( \frac{%
\eta ^3}3+\eta \sigma ^2\right) +\frac{2a\gamma }{\omega _o}\eta \sigma ,
\label{kinkamp}
\end{equation}

\begin{equation}
\frac{d\sigma }{d\tau }=-\frac{8\gamma }{3\omega _o}\eta ^2\sigma +\frac{%
2a\gamma }{3\omega _o}\eta ^2.  \label{kinkphase}
\end{equation}

\noindent In order to solve equations (\ref{kinkamp}) and (\ref{kinkphase})
we introduce the new variables

\begin{equation}
\xi \left( \tau \right) =\eta ^2\left( \tau \right) \qquad ;\qquad \kappa
\left( \tau \right) =\sigma \left( \tau \right) -\frac a4
\end{equation}

\noindent so that the system (\ref{kinkamp}), (\ref{kinkphase}) is cast into
the form

\begin{equation}
\frac{d\xi }{d\tau }=4\gamma b_1\xi -\frac{8\gamma }{3\omega _o}\xi ^2-\frac{%
8\gamma }{\omega _o}\xi \kappa ^2\qquad ;\qquad \frac{d\kappa }{d\tau }=-%
\frac{8\gamma }{3\omega _o}\xi \kappa ,  \label{system}
\end{equation}

\noindent where

\begin{equation}
b_1=\frac{a^2}{8\omega _o}-b>0.
\end{equation}
\noindent A particular solution of the system of equations (\ref{system})
can be obtained for $\kappa =0$. Thus 
\begin{equation}
\sigma =\frac a4\qquad ;\qquad \eta ^2\left( \tau \right) =3\omega _ob_1%
\frac{\eta ^2\left( 0\right) e^{4\gamma b_1\tau }}{3\omega _ob_1+2\eta
^2\left( 0\right) \left( e^{4\gamma b_1\tau }-1\right) }.
\end{equation}

\noindent Solving equation (\ref{dense}) for $\widetilde{u}$, provided $%
\widetilde{\rho }$ is given by (\ref{caviton}) and (\ref{kink}) one finds

\[
\widetilde{u}\left( x;\tau \right) =\frac{\lambda \sqrt{\sigma _v^2+3}\cosh
^2z}{2\omega _o\omega _q\left( \cosh ^2z-2\eta ^2\right) }* 
\]

\begin{equation}
\ast \left[ C+4\gamma \eta \left( b_1-\frac{2\eta ^2}{\omega _o}\right)
\tanh z+\frac{16\gamma \eta ^3}{3\omega _o}\tanh ^3z+a\frac{\eta ^2-\cosh ^2z%
}{\cosh ^2z}\right] ,
\end{equation}
\noindent where

\begin{equation}
z\left( x;\tau \right) =\eta \left( \tau \right) \left[ x-\mu \left( \tau
\right) +\beta \right] ,
\end{equation}

\begin{equation}
C=a\left[ 1-\eta ^2\left( 0\right) \right] +\frac{2\omega _o\omega _q}{%
\lambda \sqrt{\sigma _v^2+3}}\left[ 1-2\eta ^2\left( 0\right) \right] \left[
1+u_o\left( 0\right) \right] .
\end{equation}
The solutions for the mean velocity of the soliton and the corresponding
voltage amplitude are shown in Figs. 1 and 2 respectively.  We note that
the solitary wave corresponds to a self-contained droplet of charge which
separates (decelerates) from the core of the beam and approaches a fixed
separation at sufficiently long times.  The reason for this behavior is
the fact that the driving force due to the wake decays rapidly as
the soliton detunes from the resonator frequency.  At sufficient detuning, 
the wake no longer contains enough dissipation to cause further deceleration.
The resonator voltage decreases in a corresponding fashion.  It is interesting
to note that the charge contained in the soliton remains self-organized over
very long times despite the presence of dissipation.  This situation is
rather unique and is due to the peculiar character of the wake force from
the resonator.

\begin{figure}[t]
\psfig{file=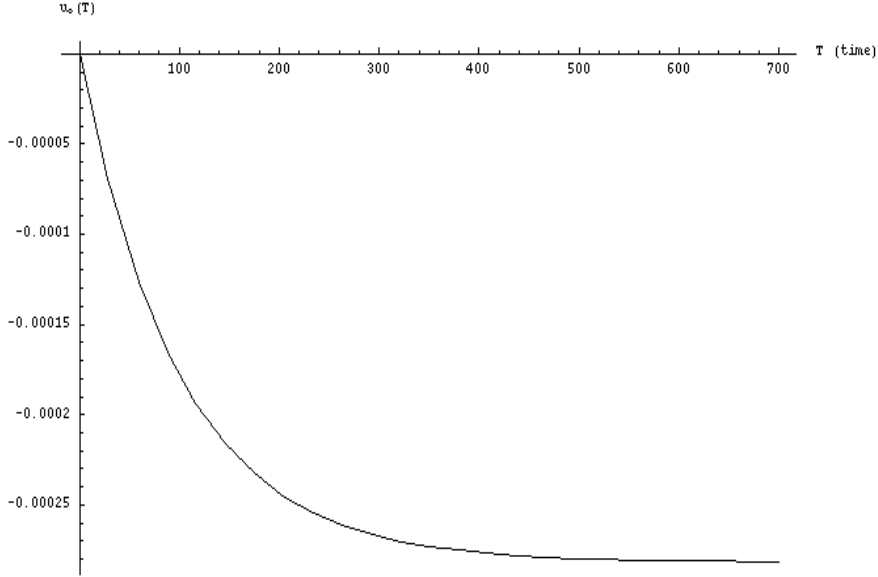,height=3.5in,width=5.0in}
\centering
\caption{ 
\label{fig:velocity}
Mean velocity of the solitary wave due to a resonator impedance.  Solitons
decelerate at first due to the dissipative part of the wakefield.  However,
over long times, they approach a steady state where the wakefields have
sufficiently decayed due to the finite resonator bandwidth.}
\end{figure}

\begin{figure}[t]
\psfig{file=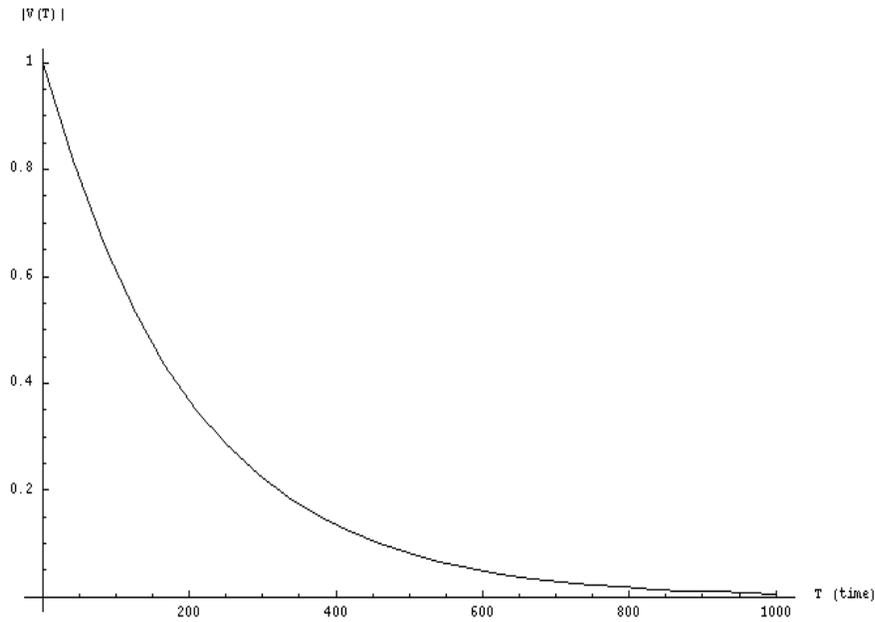,height=3.5in,width=5.0in}
\centering
\caption{ 
\label{fig:amplitude}
Voltage amplitude on the resonator.  The voltage first grows due to the
longitudinal impedance, followed by oscillations which result from the
interference of energy between the solitary waves and the core of the beam.
The envelope of the amplitude eventually decays as detuning occurs.}
\end{figure}

In Figs. 3 and 4 we show the corresponding mean velocity and voltage from
a coasting beam simulation previously reported.  The behavior is manifestly
similar to that predicted by Eq. (35), (53) and (61) , though no attempt has 
been made to check the precise scaling of the physical quantities.

\begin{figure}[t]
\psfig{file=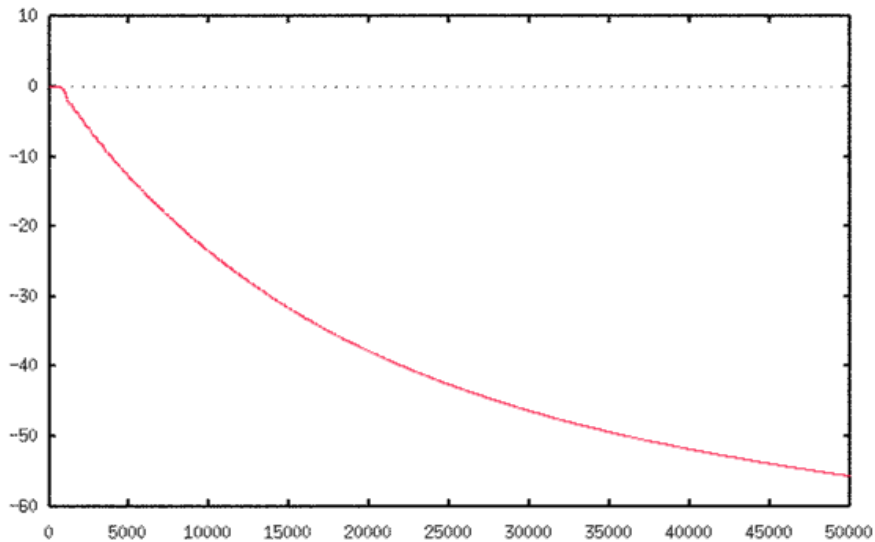,height=3.5in,width=5.0in}
\centering
\caption{ 
\label{fig:velsim}
Mean velocity of the solitary waves from the simulation showing deceleration
toward a fixed maximum energy separation.  There is good qualitative
agreement with the analytical result.}
\end{figure}

\begin{figure}[t]
\psfig{file=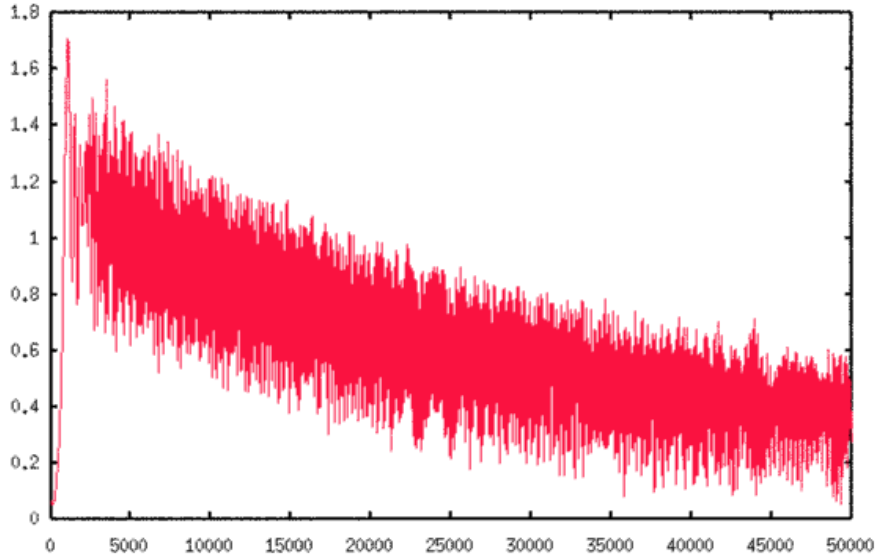,height=3.5in,width=5.0in}
\centering
\caption{ 
\label{fig:voltsim}
Voltage amplitude on the resonator from the particle simulation.
There is good qualitative agreement between the analytical results
and the voltage envelope shown.}
\end{figure}

\section{Conclusions}

In this work we have derived a set of equations for solitary waves on a
coasting beam using a renormalization group approach.  This procedure has
led to a specific set of evolution equations in the practical case of a
cavity resonator of finite Q.  The resulting set of equations can be
solved analytically under certain assumptions, and this leads to an explicit
form for the soliton and its behavior over time.  We find, in contrast to
other solitary waves in the presence of dissipation, that solitons can 
persist over long times and do so by decelerating from the core of the
beam.  This deceleration leads to detuning and the decay of the driving 
voltage.  The result is that a nearly steady state is reached, albeit with
a gradually decreasing soliton strength, but fixed maximum energy separation.
\par
Good qualitative agreement between the analytic results and the simulations 
have been observed.  We note that such a process may well indicate a method
by which well-defined droplets can occur in the halo of intense stored beams.
Further study of this problem, and the application of the RG approach to
bunched-beam evolution will be considered in future work.

\section{ACKNOWLEDGMENTS}

The authors gratefully acknowledge the continuing support of D. Finley and
S. Holmes for the pursuit of this erudite topic.  The authors also gratefully
acknowledge helpful discussions with Alejandro Aceves and Jim Ellison.

\end{document}